\documentclass[12pt,a4paper,epsf]{article}
\usepackage{graphics}
\usepackage{amssymb,amsmath}
\usepackage[dvips]{lscape,graphicx}
\usepackage{cite}
\usepackage{longtable}
\usepackage{slashed}
\usepackage{bbold}

\newcommand{\ct}{\cite}
\newcommand{\lb}{\label}

\newcommand{\bc}{\begin{center}}
\newcommand{\ec}{\end{center}}
\newcommand{\bd}{\begin{displaymath}}
\newcommand{\ed}{\end{displaymath}}
\newcommand{\be}{\begin{equation}}
\newcommand{\ee}{\end{equation}}
\newcommand{\ba}{\begin{array}}
\newcommand{\ea}{\end{array}}
\newcommand{\bea}{\begin{eqnarray}}
\newcommand{\eea}{\end{eqnarray}}
\newcommand{\bt}{\begin{tabular}}
\newcommand{\et}{\end{tabular}}

\newcommand{\bp}{\begin{picture}}
\newcommand{\ep}{\end{picture}}
\newcommand{\bfi}{\begin{figure}}
\newcommand{\efi}{\end{figure}}

\def\fun#1#2{\lower3.6pt\vbox{\baselineskip0pt\lineskip.9pt
\ialign{$\mathsurround=0pt#1\hfil##\hfil$\crcr#2\crcr\sim\crcr}}}

\begin{document}

\title{\LARGE \bf {New Resonances at LHC are possible.\\\vspace{0.5cm}
Multiple Point Principle\\ and New Bound States in the Standard Model}}
\author{\large\bf
  L.V. Laperashvili${}^{1}$\footnote{laper@itep.ru},\;
  H.B. Nielsen${}^{2}$\footnote{hbech@nbi.dk},\;
  C.D. Froggatt${}^{3}$\footnote{c.froggatt@physics.gla.ac.uk},\\
\large\bf  
  B.G. Sidharth${}^{4}$\footnote{birlasc@gmail.com},\; 
  and
  C.R. Das${}^{5}$\footnote{das@theor.jinr.ru}\\\\
{\large \it ${}^{1}$ The Institute of Theoretical and
Experimental Physics,}\\
{\large \it National Research Center ``Kurchatov Institute'',}\\
{\large \it Bolshaya Cheremushkinskaya, 25, 117218 Moscow, Russia}\\\\
{\large \it ${}^{2}$ Niels Bohr Institute,}\\
{\large \it Blegdamsvej, 17-21, DK 2100 Copenhagen, Denmark}\\\\
{\large \it ${}^{3}$ Department of Physics,}\\
{\large \it Glasgow University, UK}\\\\
{\large \it ${}^{4}$ International Institute of Applicable Mathematics}\\
{\large \it and Information Sciences,}\\
{\large \it B.M. Birla Science Centre}\\
{\large \it Adarsh Nagar, 500063 Hyderabad, India}\\\\
{\large \it ${}^{5}$ Bogoliubov Laboratory of Theoretical Physics}\\
{\large \it Joint Institute for Nuclear Research}\\
{\large \it International Intergovernmental Organization,}\\
{\large \it Joliot-Curie 6, 141980 Dubna, Moscow region, Russia}}

\date{}
\maketitle

\thispagestyle{empty}

\vspace{0.5cm}



\thispagestyle{empty}

\clearpage \newpage

\begin{abstract}

In the present paper we argue that the correction to the Higgs mass coming from
the bound state of 6 top and 6 anti-top quarks, predicted early by C.D. Froggatt,
H.B. Nielsen and L.V. Laperashvili, leads to the Standard Model (SM) vacuum stability
and confirms the accuracy of the multiple point principle (principle of degenerate vacua)
for all experimentally valued SM parameters (Higgs mass, top-quark mass, etc.).
The aim to get the vacua degeneracy requires a mass of the bound state in the region of 770 GeV.

\end{abstract}

\newpage

\tableofcontents

\newpage

\section{Multiple Point Principle (MPP)}

In this investigation we assumed a new law of Nature named as a
Multiple Point Principle (MPP), which was suggested first by D.L. Bennett and
H.B. Nielsen in Ref. \ct{1}. MPP postulates:\\
{\it There exist in Nature several degenerate vacua with very small
energy density, or cosmological constants (CC).}
This principle is based on the discovery that a cosmological constant of our
Universe is extremely small, almost zero \ct{2,3,4}.

The MPP theory was developed in a lot of papers by H.B. Nielsen, D.L. Bennett, C.D. Froggatt,
R.B. Nevzorov, L.V. Laperashvili, C.R. Das (see for example, Refs.
\ct{5,6,7,8,9,10,11,12,13,14,15,16,17,18,19,20,21,22}) and recently by other authors \ct{23,24,25,26}.

Vacuum energy density of our Universe is the Dark Energy (DE), which is related with
cosmological constant $\Lambda$ by the following way:
\be
\rho_{DE} = \rho_{vac} = (M^{red}_{Pl})^2\Lambda.  \lb{1} \ee
Here $M^{red}_{Pl}$ is the reduced Planck mass: $M^{red}_{Pl}\simeq 2.43 \times 10^{18}$ GeV.

Recent cosmological measurements (see Ref. \ct{27}) give:
\be  \rho_{DE} \simeq (2\times 10^{-3}\; {\rm{ eV}})^4.  \lb{2} \ee
According to (\ref{2}), we have a tiny value of CC:
\be \Lambda \simeq 10^{-84}\; {\rm GeV}^4. \lb{3} \ee
This tiny value of $\rho_{DE}$ was first predicted by B.G. Sidharth in 1997 year \ct{2,3}.
In 1998 year S. Perlmutter, B. Schmidt and A. Riess \ct{4} were awarded by the Nobel Prize
for discovery of the Universe accelerating expansion.

Considering extremely small cosmological constant of our Universe, Bennett, Froggatt and Nielsen
assumed only zero, or almost zero, cosmological constants for all vacua existing in the Universe.

Restricted ourselves to {\bf the pure Standard Model (SM)} we have only three vacua:
\begin{enumerate}
\item {\bf Present Electroweak vacuum}, in which we live.\\
It has vacuum expectation value (VEV) of the Higgs field equal to:
\be v_1 = v = \langle \phi_H\rangle  \approx 246 \; {\rm{GeV}}.  \lb{4} \ee
\item {\bf High Higgs field vacuum} - Planck scale vacuum, which has the following VEV:
\be v_2 = v = \langle \phi_H\rangle  \sim 10^{18} \; {\rm{GeV}}.  \lb{5} \ee
\item {\bf Condensate vacuum.} This third vacuum is a very speculative possible state
inside the pure SM, which contains a lot of strongly bound states, each
bound from 6 top + 6 anti-top quarks (see Refs. \ct{28,29,30,31,32}).
\end{enumerate}

From experimental results for these three vacua, cosmological
constants - minimum of the Higgs effective potentials $V_{eff} (\phi_H)$ - are
not exactly equal to zero. Nevertheless, they are extremely small.
By this reason, Bennett, Froggatt and Nielsen assumed to consider zero cosmological
constants as a good approximation. Then according to the MPP, we have a model of the pure SM
being finetuned in such a way that these three vacua proposed have just zero energy density.

If the effective potential has three degenerate minima, then the
following requirements are satisfied:
\be
V_{eff} (\phi^2_{min1}) =  V_{eff} (\phi^2_{min2}) =  V_{eff} (\phi^2_{min3}) = 0,  \lb{6}
\ee
and
\be
V'_{eff} (\phi^2_{min1}) =  V'_{eff} (\phi^2_{min2}) =  V'_{eff} (\phi^2_{min3}) = 0, \lb{7}
\ee
where
\be
V'(\phi^2) = \frac{\partial V}{\partial \phi^2}. \lb{8}
\ee
Here we assume that:
\be
V_{eff} (\phi^2_{min1}) = V_{present}, \quad V_{eff} (\phi^2_{min2}) = V_{high-field}, \quad
{\rm{and}} \quad V_{eff} (\phi^2_{min3}) = V_{condensate}.     \lb{9}
\ee

As a result, Multiple Point Principle postulates:
{\bf There are three vacua in the SM with the same energy density, or
cosmological constant, and all cosmological constants are zero, or
approximately zero.}
Our MPP is really just an assumption about the coupling constants,
i.e. MPP provides three restrictions between the parameters of the SM,
from which zero energy density in all three vacua follows.

Here we wrote explicitly the following parameters of the SM:
\begin{itemize}
\item[] $\Lambda_{CC}$: The cosmological constant,
\item[] $g_t$ : The top Yukawa coupling,
\item[] $m^2_H$: Higgs mass squared,
\item[] $\Lambda_{QCD}$: The scale parameter of QCD.
\end{itemize}
$V_{present}$ , $V_{high-field}$ and $V_{condensate}$ are the vacuum energy
densities for these three speculated vacua.
Taking into account the experimental values for all SM parameters, we could look at
them, that $V_{present} = 0$ fixes the cosmological constant $\Lambda_{CC}$ to be
essentially zero (indeed, it is very small).
Then $V_{high-field}$ (meaning the energy density of the vacuum having the
very high Higgs field $\phi_H \sim 10^{18}$ GeV) could be taken to predict the
Higgs mass. And the $V_{condensate} = 0$ helps to predict the $g_t$ Yukawa coupling.

Assuming the existence of the two degenerate vacua in the SM:
\begin{enumerate}
\item the first Electroweak vacuum at $v_1 = v = 246$ GeV, and
\item the second Planck scale vacuum at $v_2 \sim 10^{18}$ GeV,
\end{enumerate}
Froggatt and Nielsen predicted in Ref. \ct{33} the top-quark and Higgs boson masses:
\be
M_t = 173 \pm 5\; {\rm { GeV}}; \quad M_H = 135 \pm 10\; {\rm { GeV}}. \lb{10}
\ee
In Fig. 1 it is shown the existence of the second (non-standard) minimum of the effective
potential in the pure SM at the Planck scale.

\section{New bound states (NBS) of 6 top + 6 anti-top quarks}

We assume that the vacuum, which we call ``condensate vacuum'',
is filled by bound states 6 top + 6 anti-top (see Fig. 2).
These bound states cannot be calculated perturbatively.
Our aim was to estimate/calculate the value of the mass of the bound
state of 6 top and 6 anti-top quarks speculated to exist in our model.

In Refs. \ct{28,29,30} there was suggested the existence of new bound states (NBS)
of 6 top + 6 anti-top quarks as so strongly bound systems that they
effectively function as elementary particles.
There was first assumed that
\begin{enumerate}
\item there exists $1S$-bound state 6 t + 6 anti-t - scalar particle and color
singlet;
\item that the forces responsible for the formation of new bound states (NBS)
originate from the virtual exchanges of the Higgs bosons between
top(anti-top)-quarks;
\item that these forces are so strong that they almost compensate the
mass of 12 top-anti-top quarks contained in these bound states.
\end{enumerate}

The explanation of stability of the bound state 6 t + 6 anti-t is given
by the Pauli principle:
top-quark has two spin and three color degrees of freedom (total 6).
By this reason, 6 quarks have the maximal binding energy, and 6 pairs
of $t\bar t$ in $1S$-wave state create a long lived (almost stable) colorless
scalar bound state $S$. One could even suspect that not only this most
strongly bound state $S$ of 6 t + 6 anti-t, but also some excited states
exist. It is obvious that excited to a $2S$ or $2P$ state (in the atomic physics
terminology), scalar and vector particles correspond to the more heavy
bound states of the 6 t + 6 anti-t, etc. Also there exists a new bound
state 6 t + 5 anti-t, which is a fermion similar to the quark of the 4th
generation having quantum numbers of top-quark.
These bound states are held together by exchange of the Higgs and
gluons between the top-quarks and anti-top-quarks as well as between
top and top and between anti-top and anti-top.
The Higgs field causes attraction between quark and quark as well as
between quark and anti-quark and between anti-quark and anti-quark,
so the more particles and/or anti-particles are being put together the
stronger they are bound. But now for fermions as top-quarks, the Pauli principle
prevents too many constituents being possible in the lowest state of a Bohr atom
constructed from different top-quarks or anti-top-quarks surrounding
(analogous to the electrons in the atom) the ``whole system'', analogous
to the nucleus in the Bohr atom.

Because the quark has three color states and two spin states meaning
six $(3\times 2 = 6)$ internal states, there is in fact a shell (as in the nuclear
physics) with six top quarks and similarly one for six anti-top quarks.
Then we imagine that in the most strongly bound state just this shell is
filled and closed for both top and anti-top. Like in nuclear physics
where the closed shell nuclei are the strongest bound, we consider this
NBS 6 top + 6 anti-top as our favorite candidate for the most strongly
bound and thus the lightest bound state $S$.
As a result, we expect that our bound state $S$ is appreciably lighter than its
natural scale of 12 times the top mass, which is about 2 TeV. So the
mass of our NBS $S$ should be small compared to 2 TeV.

Using our MPP we got three a priori different mass predictions for our
new bound state (see Refs. \ct{31,32}):
\begin{enumerate}
\item High-field fit: The MPP-requirement that the ``present vacuum'' is
degenerate with the ``high-field vacuum'' gives fitting mass $m_S$ of
the bound state $S$, approximately equal from 700 GeV to 800 GeV.
\item Condensate vacuum fit: We fit the mass $m_S$ in a region filled with
particles 6 t + 6 anti-t, having the lowest energy density just the
same (zero) energy density as the ``present vacuum''. With a
simple but accidentally almost true assumption, we fit the mass
$m_S$ (from condensate fit) $\approx 4m_t = 692 \; {\rm{GeV}} \pm 100\;  {\rm{GeV}}$.
\item Bag-model fit: We also make a bag-model-like crude ansatz for
the bound state of the 6 top + 6 anti-top, and seek the minimum
energy by varying bag radius $R$. With very crude inclusion of
various corrections, we reach the mass estimate
$m_S \approx 5m_t = 865 \; {\rm{GeV}} \pm 200 \;  {\rm{GeV}}$.
\end{enumerate}

Thus, we have suggested in our model of pure SM the existence of
three degenerate vacua: ``present'', ``high-field'' and ``condensate''
vacua. If several vacua are degenerate, then the phase diagram of theory
contains a special point - {\bf the Multiple Critical Point (MCP)}, at which
the corresponding phases meet together.

Here it is useful to remind you {\bf a triple point} of water analogy.
It is well known in the thermal physics that in the range of fixed
extensive quantities: volume, energy and a number of moles, the
degenerate phases of water (namely, ice, water and vapor)
exist on the phase diagram (P, T) shown by Fig. 3, where we have
pressure P and temperature T. Fig. 3 gives the critical (triple) point $O$ with:
\be T_c \approx 0.01^0 {\rm C},\quad P_c \approx 4.58\;  {\rm mm\;  Hg}. \lb{11} \ee
This is a triple point of water analogy.

The idea of the Multiple Point Principle has its origin from the lattice
investigations of gauge theories. In particular, Monte Carlo simulations
of $U(1)-$, $SU(2)-$ and $SU(3)-$ gauge theories on lattice indicate the
existence of the triple critical point.

\section{Two-loop corrections to the Higgs mass from the
effective potential}

The prediction (\ref{10}) by Froggatt and Nielsen for the mass of the Higgs boson
was improved in Ref. \ct{34} by calculations of the two-loop radiative corrections
to the effective Higgs potential. The prediction of Ref.\ct{34}: $M_H = 129 \pm 2$ GeV provided
the possibility of the theoretical explanation of the value $M_H \simeq 125.7$ GeV
observed at LHC.

The authors of Ref. \ct{35} extrapolated the SM parameters up to the high (Planck)
energies with full 3-loop NNLO RGE precision.
From Degrassi et al. calculation \ct{34}, the effective Higgs field potential
$V_{eff}(\phi_H)$ has a minimum, which goes slightly under zero, so that the
present vacuum is unstable for the experimental Higgs mass $125.09 \pm 0.24$ GeV,
while the value that would have made the second minimum just degenerate with the
present vacuum energy density would be rather $m_H \simeq 129.4$ GeV.

\section{Higgs mass and vacuum stability/metastability in the
Standard Model}

A theory of a single scalar field is given by the effective potential $V_{eff}(\phi_c)$, which is a
function of the classical field $\phi_c$. In the loop expansion $V_{eff}$ is given by
\be
V_{eff} = V (0) + \Sigma_{n=1}V^{(n)},    \lb{12}   \ee
where $V (0)$ is the tree-level potential of the SM:
\be
V (0) = - \frac 12 m^2\phi^2 + \frac 14 \lambda \phi^4.   \lb{13} \ee
The vast majority of the available experimental data is consistent with
the SM predictions. No sign of new physics has been detected. Until
now there is no evidence for the existence of any particles other than
those of the SM, or bound states composed of other particles. All
accelerator physics seems to fit well with the SM, except for neutrino
oscillations. These results caused a keen interest in possibility of
emergence of new physics only at very high (Planck scale) energies,
and generated a great attention to the problem of the vacuum stability:
whether the electroweak vacuum is stable, unstable, or metastable.
A largely explored scenario assumes that new physics interactions
only appear at the Planck scale $M_{Pl} = 1.22 \times 10^{19}$ GeV. According to
this scenario, we need the knowledge of the Higgs effective potential
$V_{eff}(\phi)$ up to very high values of $\phi$.

The loop corrections lead the $V_{eff}$ to values of $\phi$, which are much
larger than $v_1\approx 246$ GeV - the location of the electroweak (EW) minimum.
The effective Higgs potential develops a new minimum at $v_2 \gg v$. The position
of the second minimum depends on the SM parameters, especially on the top and Higgs
masses, $M_t$ and $M_H$.  This $V_{eff}$ can be higher or lower than the EW one showing
a stable EW vacuum (in the first case), or metastable one (in the second case).

Then considering the lifetime $\tau$ of the false vacuum (see Ref. \ct{36})
and comparing it with the age of the Universe $T_U$, we see that, if $\tau$ is
larger than $T_U$, then our Universe will be sitting on the metastable false
vacuum, and we deal with the scenario of metastability.

Usually the stability analysis is presented by stability diagram in the
plane $(M_H; M_t)$. This plot is shown in Fig. 4, where we see that the plane
$(M_H; M_t)$ is divided into three different sectors:
\begin{enumerate}
\item An absolute stability (cyan) region, where $V_{eff} (v_1) < V_{eff} (v_2)$,
\item a metastability (yellow) region, where $V_{eff} (v_2) < V_{eff} (v_1)$, but $\tau > T_U$,
and
\item an instability (green) region, where $V_{eff} (v_2) < V_{eff}(v_1)$ and $\tau < T_U$.
The stability line separates the stability and the metastability regions
and corresponds to $M_t$ and $M_H$ obeying the condition $V_{eff} (v_1) = V_{eff} (v_2)$.
The instability line separates the metastability and the instability
regions. It corresponds to $M_t$ and $M_H$ for $\tau = T_U$.
In Fig. 4 the black dot indicates current experimental values
$M_H \approx 125.7$ GeV and $M_t \approx 173.34$ GeV (see Data Group \ct{27}).
It lies inside the metastability region, and could reach and even cross the stability line
within $3\sigma$. The ellipses take into account $1\sigma; 2\sigma; 3\sigma;$ according to
the current experimental errors.
\end{enumerate}

When the black dot sits on the stability line, then this case is named
``critical'', according to the MPP concept: the running quartic coupling
and the corresponding beta-function vanish at the Planck scale:
\be   \lambda_{eff}(\phi_0) \simeq 0,\quad \beta(\lambda_{eff}(\phi_0)) \simeq 0.  \lb{14} \ee
Fig. 4 shows that the black dot, existing in the metastability
region, is close to the stability line, and the ``near-criticality'' can be
considered as the most important information about the Higgs boson \ct{35}.

\subsection{Two-loop corrections}

From Degrassi et al.'s calculation \ct{34} of the effective Higgs field potential
$V_{eff}(\phi_H)$, there is a minimum in this potential, but it goes slightly under
zero, so that the present vacuum is unstable for the experimental Higgs
mass $125.7 \pm 0.24$ GeV, while the value that would have made the second minimum
just degenerate with the present vacuum energy density would be rather $m_H$
approximately equal to 129.4 GeV (from MPP by Degrassi et al. \ct{34}).

Still neglecting new physics interactions at the Planck scale, we can
consider for large values of $\phi$:
\be    V_{eff} (\phi) \simeq \frac 14 \lambda_{eff}(\phi)\phi^4,  \lb{15} \ee
where $\lambda_{eff}(\phi)$ depends on $\phi$ as the running quartic coupling
$\lambda(\mu)$ depends on the running energy scale $\mu$. Then we have the one-loop,
two-loops, three-loops, etc. expressions for $V_{eff}$ .

The corresponding up to date Next-to-Next-to-Leading-Order (NNLO)
results were published in several references, and show that for a large
range of values of $M_H$ and $M_t$  the Higgs effective potential has a
minimum.

In Fig. 5 blue lines (thick and dashed) present the RG evolution of $\lambda(\mu)$
for current experimental values $M_H \simeq 125.7$ GeV and $M_t\simeq 173.34$ GeV.
The thick blue line corresponds to the central value of $\alpha_s = 0.1184$ and
dashed blue lines correspond to its errors equal to $\pm 0.0007$.
The red solid line of Fig. 5 shows the running of the $\lambda_{eff} (\phi)$ for
$M_H \simeq 125.7$ GeV and $M_t \simeq 171.43$ GeV, which just corresponds to the
stability line, that is, to the stable vacuum.
In this case the minimum of the $V_{eff} (\phi)$ exists at the
$\phi = \phi_0 \approx 2.22 \times 10^{18}$ GeV, where according to MPP:
$\lambda_{eff} (\phi_0) = \beta(\lambda_{eff}(\phi_0)) = 0$.
But as it was shown in Refs. \ct{34,35} by Degrassi et al. and Buttazzo et al. this
case does not correspond to the current experimental values.

According to Ref. \ct{34}, the stability line shown in Fig. 5 by the red thick line
corresponds to $M_H = 129.4 \pm 1.8$ GeV.
We see that the current experimental values of $M_H$ and $M_t$ show the
metastability of the present Universe vacuum and MPP law is not exact.

\section{Could the MPP be exact due to corrections from the
new bound state 6 t + 6 anti-t ?}

In our Ref. \ct{22} it was shown that if we take seriously the correction to the Higgs
mass coming from the bound state $S$ of 6 top and 6 anti-top quarks,
predicted early by C.D. Froggatt, H.B. Nielsen and L.V. Laperashvili \ct{28,29,30}, then
we can obtain the SM vacuum stability and confirm the accuracy of the MPP for all
experimentally valued parameters (Higgs mass, top-quark mass, etc.).

The relation between $\lambda$ and the Higgs mass is:
\be \lambda(\mu) = \frac{G_F}{\sqrt 2}M_H^2 + \Delta\lambda(\mu), \lb{16}  \ee
where $G_F$ is the Fermi coupling, and the term $\Delta\lambda(\mu)$
denotes corrections arising beyond the tree level case (many loops contribution).

Estimating different contributions of the bound state $S$,
we have found that the main Feynman diagram, correcting the effective
Higgs self-interaction coupling constant $\lambda(\mu)$ by the constant $\lambda_S$,
is the diagram shown in Fig. 6, which contains the bound state $S$ in the loop. And
now we have:
\be \lambda(\mu) = \frac{G_F}{\sqrt 2}M_H^2 + \Delta\lambda(\mu) + \delta\lambda(\mu), \lb{17}  \ee
where the term $\delta\lambda(\mu)$ denotes the loop corrections
to the Higgs mass arising from all NBS. The main contribution to $\delta\lambda(\mu)$
is the term $\lambda_S$, which corresponds to the contribution of the diagram shown in Fig. 6:
\be   \delta\lambda(\mu) = \lambda_S + ... \lb{18}  \ee
Defining a quantity $b$, which denotes the radius of the bound state $S$
measured with top quark Compton wave length $1/m_t$ as unit by
equations:
\be     \langle {\vec r}^2\rangle  = 3r^2, \quad r_0 = \frac {b}{m_t}, \lb{19}  \ee
we obtained the dominant diagram correction contribution given by Fig. 6:
\be \lambda_S \approx \frac 1{\pi^2} \left(\frac{6g_t m_t}{b m_S}\right)^4
\approx 0.01,  \lb{20}  \ee
where we have used the estimated or measured values:
$$g_t = 0.935,\quad m_t = 173\;  {\rm{GeV}},\quad m_S \approx 750\; {\rm{GeV}},
\quad b \approx 2.34\; {\rm{or}}\;  2.43.$$
Here we assumed the existence of the resonance with mass $M_F \simeq 750$
GeV, which was supposed by the LHC measurements \ct{22}.

Using the rather small deviation from the perfect MPP, obtained earlier by
Degrassi et al. \ct{34}:
\be \lambda_{high-field} = - 0.01 \pm 0.002, \lb{21}  \ee
and requiring to be cancelled it by the correction from the bound state
$S$, we got the following result:
\be   \lambda_S = 0.01 \pm 0.002. \lb{22}  \ee
This contribution compensates the asymptotic value of $\lambda_{asym} = - 0.01$,
which was earlier obtained by Degrassi et al. \ct{34}, and therefore
transforms the metastability of the EW vacuum into the stability,
what confirms the MPP concept.

Unfortunately, now recently much discussed at LHC statistical
fluctuation peak $F(750)$ has been revealed to be just a fluctuation
(see Refs. \ct{37,38,39,40}). It is really sad for our picture that
the enhancement known as $F(750)$ decaying into the two gammas and having
just the right mass, confirming the vacuum stability, were washed out
at LHC so that no statistics remains supporting it.
In this connection, we suggest that the bound state $S$ should really exist
(see recent paper by Holger Nielsen \ct{32} and Ref. \ct{31}).
Very accidentally its mass coincides with the mass of this fluctuation
$F(750)$ GeV.
In the paper of Ref. \ct{32} Holger Nielsen gives an estimation of
the mass $m_S$ of the bound state $S$ by three different ways:
\begin{enumerate}
\item from Bag Model: $m_S \approx 830$ GeV;
\item assuming $V_{present} = V_{condensate}$: $m_S \approx 690$ GeV;
\item and $V_{present} = V_{high-field}$ : $m_S \approx 780$ GeV.\\
Average gives the result: $m_S \approx 770$ GeV.
\end{enumerate}

Remarkable that these three a priori quite different ways give an
almost the same results. It means that the degeneracies of the vacua
as implicated by MPP could be claimed to have been tested by direct
calculations using the parameters of the pure SM.
If we can perform non-perturbative calculations sufficiently accurately,
we should be able to calculate the above postulated vacua.
The ``high-field'' vacuum requires that the SM is valid up to about $10^{18}$
GeV. One could imagine that a new physics might modify our calculations,
but LHC has already put severe limits, telling that there is no new physics
up to a scale of the order of 1 TeV. Then our proposed bound state of a mass
of the order of 770 GeV is expected to be not very sensitive to at present
acceptable new physics.

At the end, we would like to emphasize: Since our picture is {\bf PURE STANDARD MODEL},
everything can be calculated, in principle. It is only a question of better techniques:
Bethe-Salpeter equation, or lattice theory with Higgs field on the lattice.
What gives a more accurate check of the MPP and calculation of the bound
state mass? This is just a work for theorists.
If the answer is negative, then it would mean that the Standard Model
is not right also non-perturbatively.

\section{Conclusions}

In this paper we considered that the correction to the Higgs mass coming from
the bound state of 6 top and 6 anti-top quarks leads to the Standard Model vacuum
stability for all experimentally valued parameters (Higgs mass, top-quark mass, etc.).

We assumed the existence of a new law of Nature, which was named as a
Multiple Point Principle (MPP). The MPP postulates: {\it There are three vacua in the SM
with the same energy density, or cosmological constant, and all cosmological constants
are zero, or approximately zero.} We considered the following three
degenerate vacua in the SM: a) the first Electroweak vacuum at $v_1 = v = 246$ GeV,
b) the second Planck scale vacuum at $v_2 \sim 10^{18}$ GeV, and the third
Condensate vacuum, which contains a lot of strongly bound states, each from
6 top + 6 anti-top quarks.

We assumed that there exists a new resonance with mass
$m_S\approx 770$ GeV, which is a new scalar $S$ bound state 6t+6anti-t,
earlier predicted by C.D. Froggatt, H.B. Nielsen and L.V. Laperashvili.
It was shown that this resonance can provide the vacuum stability and exact accuracy
of the Multiple Point Principle.
We calculated the main contribution of the $S$-resonance to the renormalization group
evolution of the Higgs quartic coupling $\lambda$, and showed that the resonance with mass
$m_S \approx 750$ GeV, having the radius $r_0 = b/m_t$ with $b \approx 2.34$, gives
the positive contribution to $\lambda(\mu)$ equal to the $\lambda_S \approx +0.01$.
This contribution compensates the asymptotic value of the $\lambda \approx - 0.01$,
which was earlier obtained in Ref. \ct{34}, and therefore transforms the metastability
of the EW vacuum into the stability.

We predict that LHC can find a new resonance at energy $\sim 1$ TeV.

\section*{Acknowledgments}
L.V.L. greatly thanks the B.M. Birla Science Centre (Hyderabad, India) and personally Prof.
B.G. Sidharth, for hospitality, collaboration and financial support. H.B.N. wishes to
thank the Niels Bohr Institute for the status of professor emeritus and corresponding
support. C.R.D. is thankful to Prof. D.I. Kazakov for support.

\newpage

\begin{figure}
\centering
\includegraphics[scale=0.55]{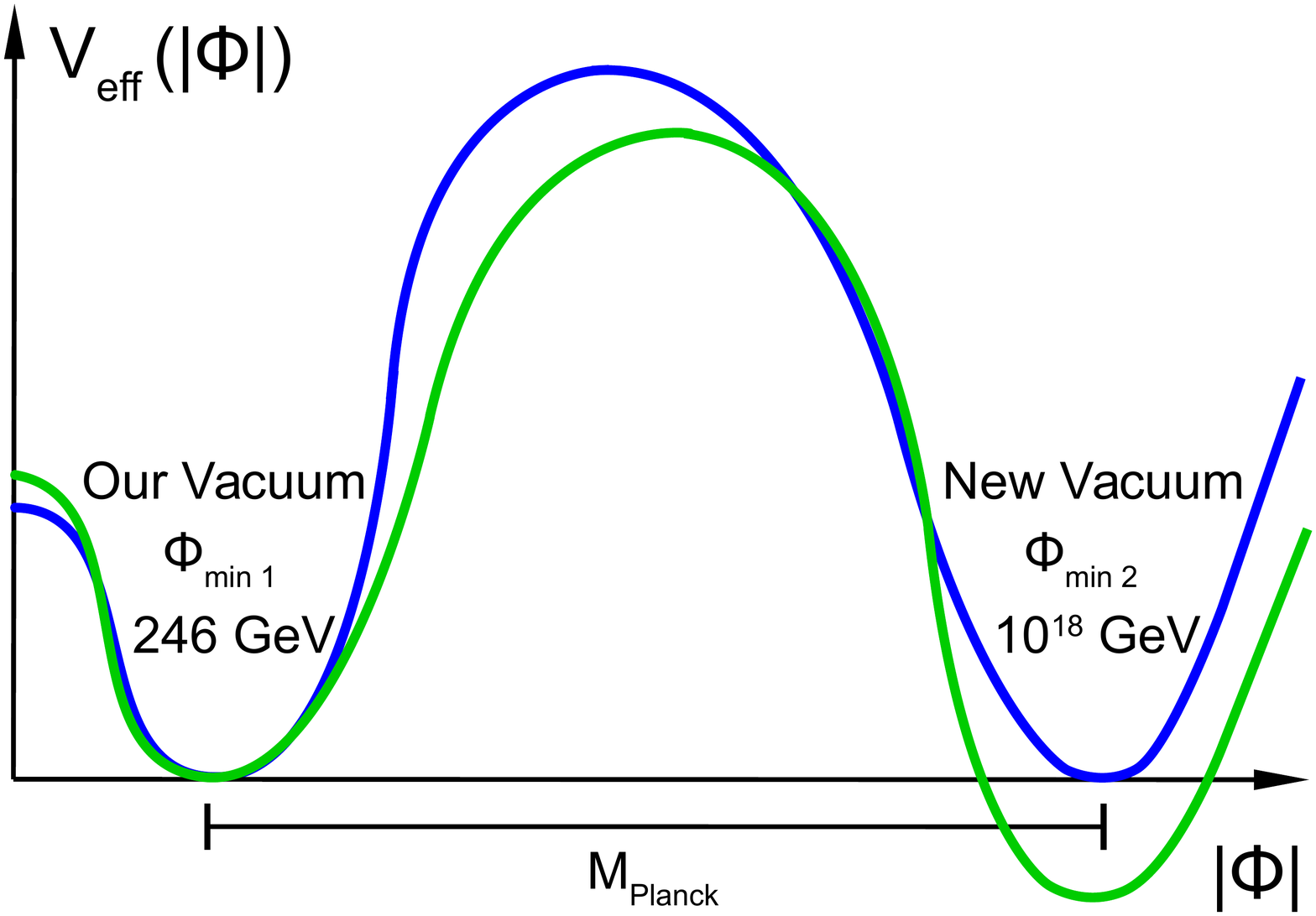}
\caption
{The existence of the second (non-standard) minimum
of the effective potential of the pure SM at the Planck scale.}
\end{figure}

\newpage

\begin{figure}
\centering
\includegraphics[scale=0.55]{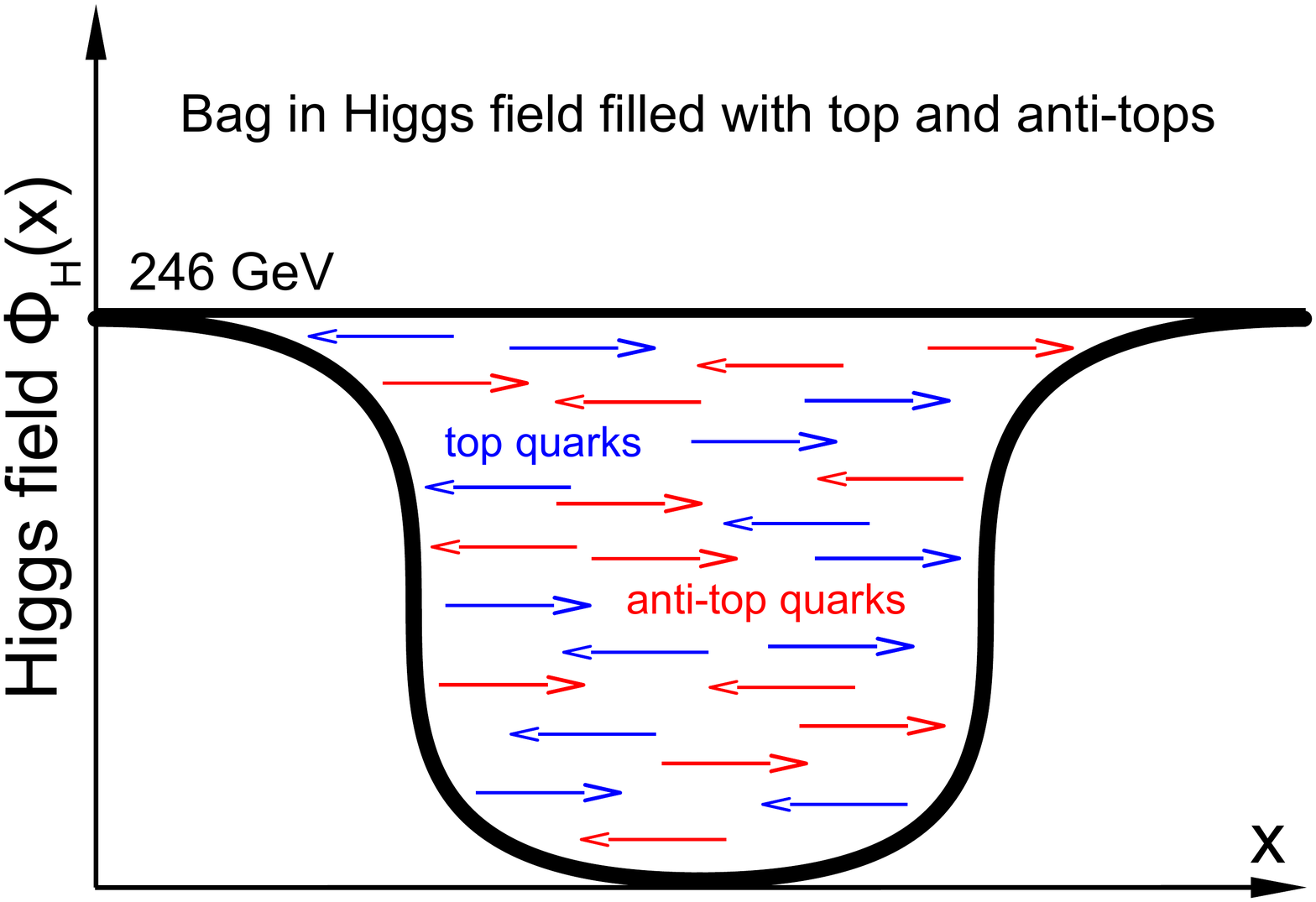}
\caption
{Condensate vacuum filled by bound states 6 top + 6 anti-top.}
\end{figure}

\newpage

\begin{figure}
\centering
\includegraphics[scale=0.95]{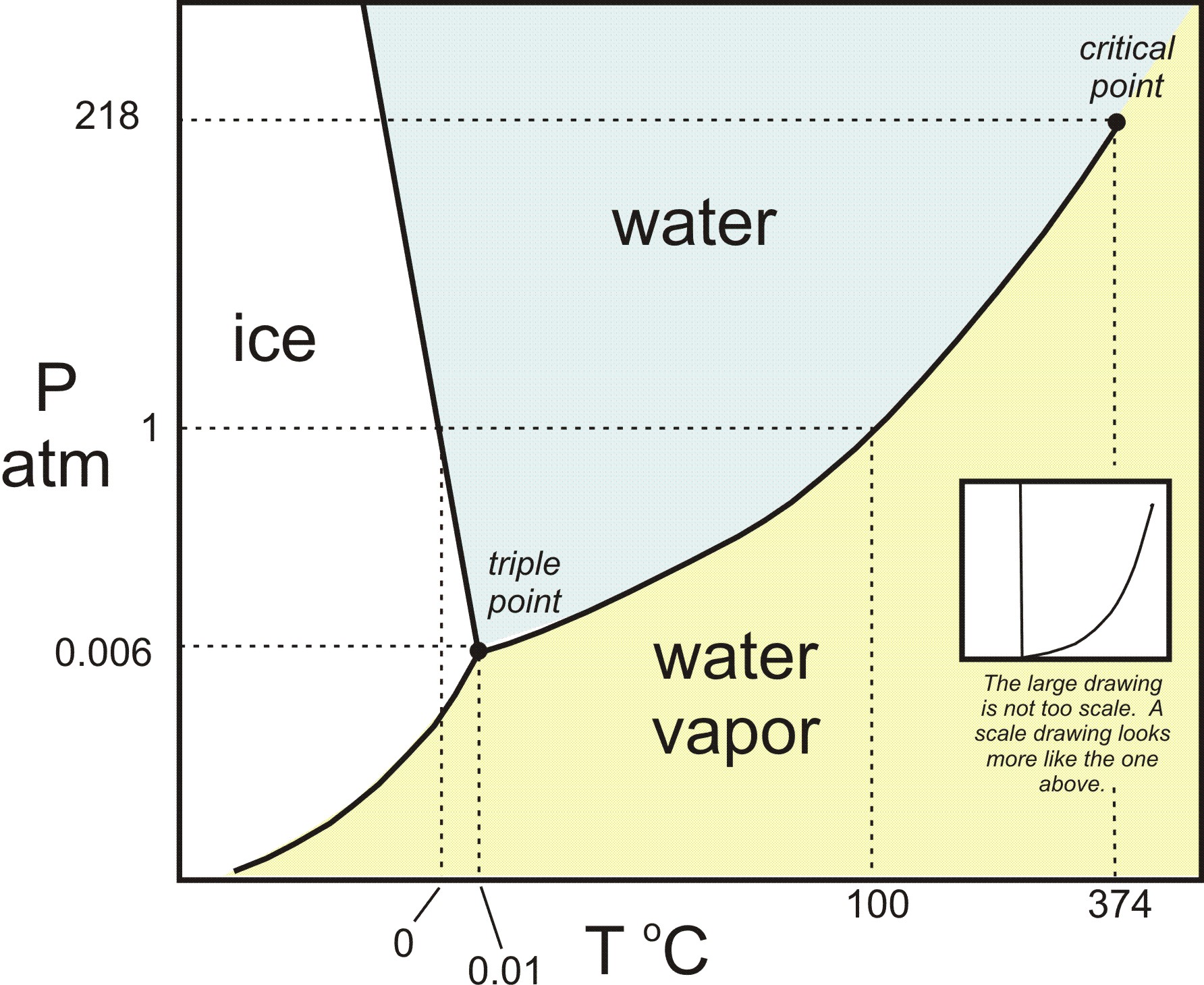}
\caption
{Phase diagram with a triple point O of water analogy.}
\end{figure}

\newpage

\begin{figure}
\centering
\includegraphics[scale=0.65]{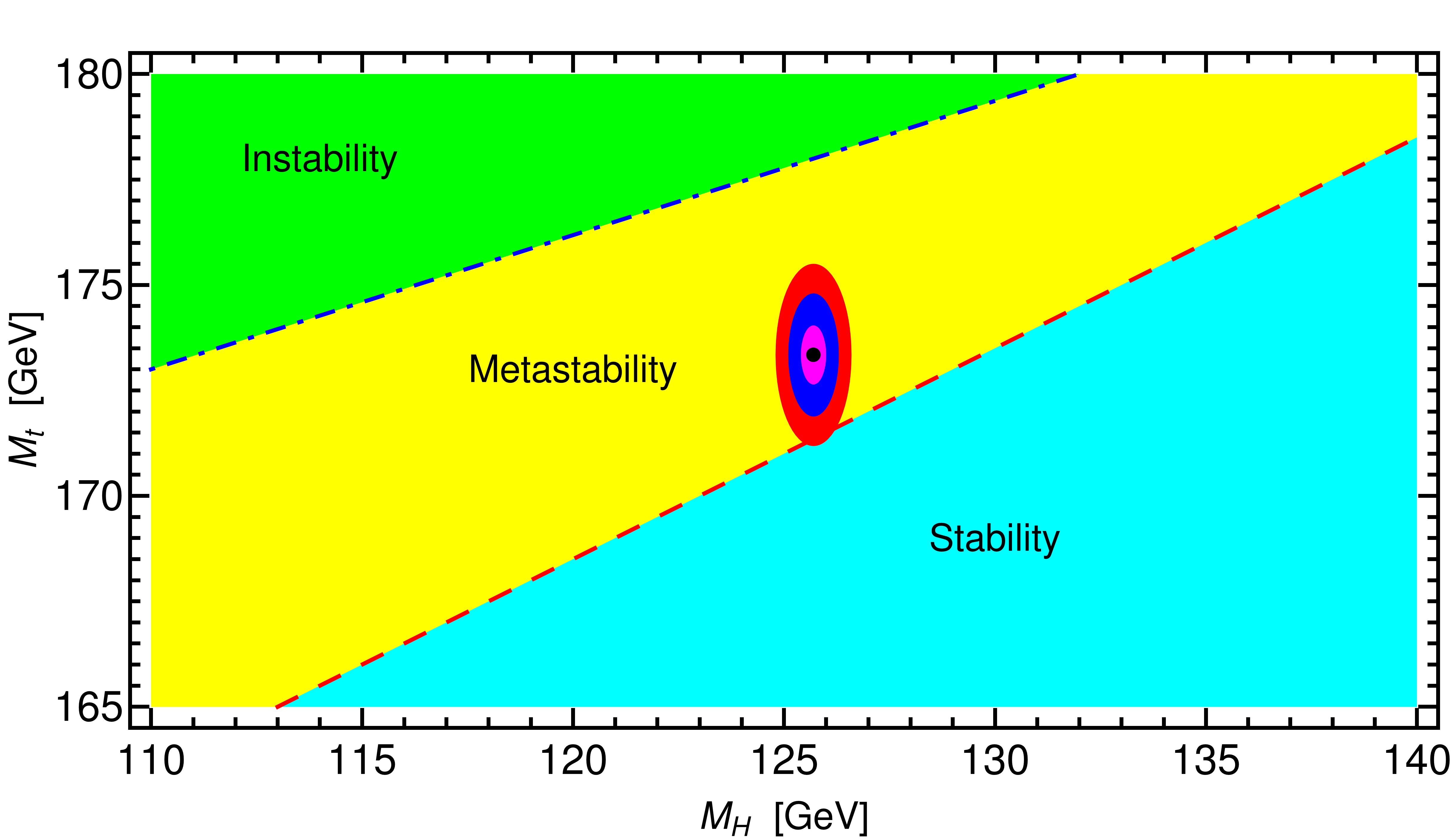}
\caption
{The stability phase diagram obtained according to the standard analysis. The
$(M_H, M_t)$ plane is divided in three sectors: absolute stability, metastability and instability
regions. The dot indicates current experimental values $M_H \simeq 125.7$ GeV and $M_t \simeq 173.34$
GeV. The ellipses take into account $1\sigma, 2\sigma$ and $3\sigma$, according to the current
experimental errors.}
\end{figure}

\newpage

\begin{figure}
\centering
\includegraphics[scale=0.65]{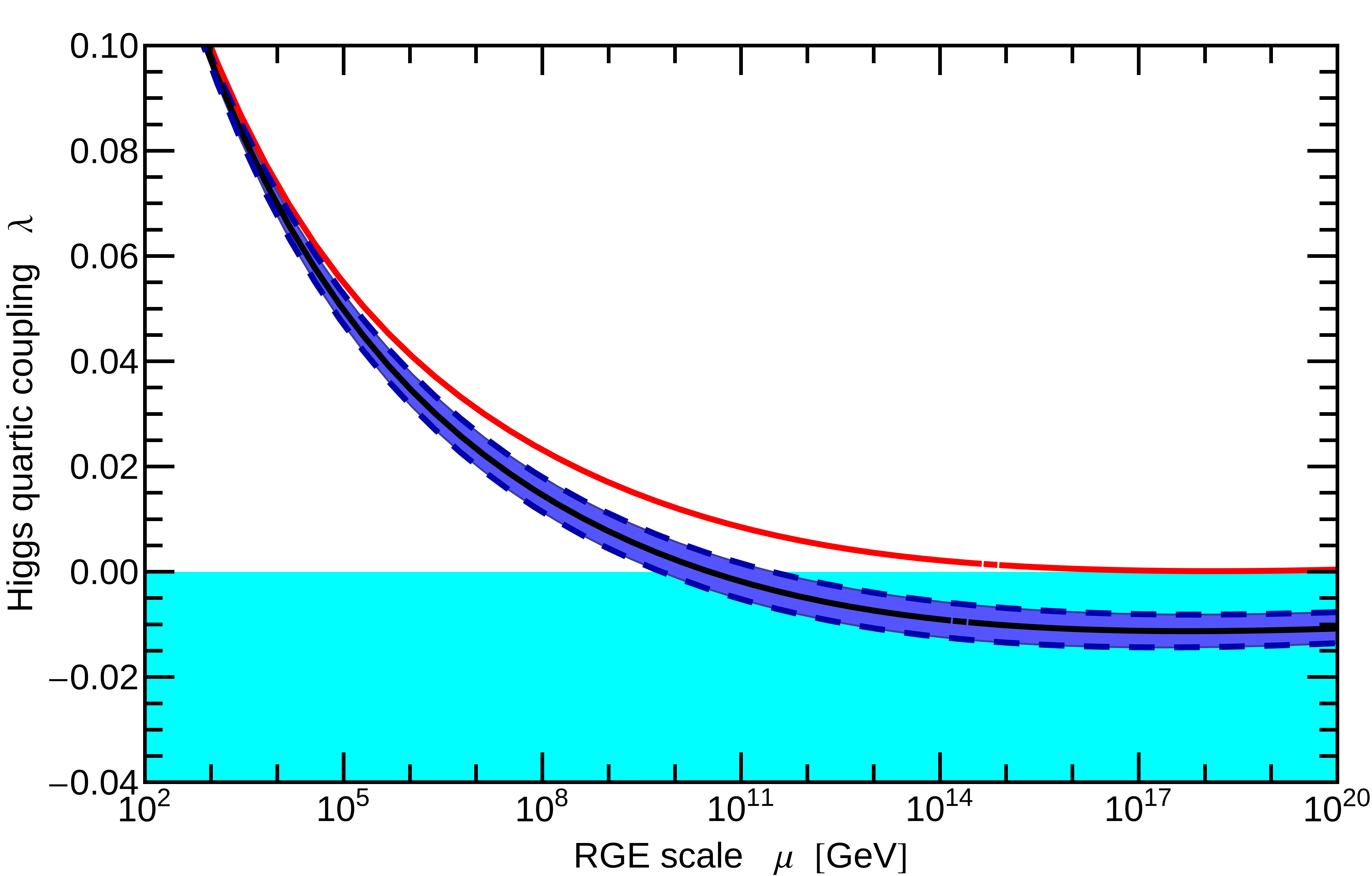}
\caption
{The RG evolution of the Higgs self-coupling $\lambda$ for $M_t \simeq 173.34$ GeV
and $\alpha_s = 0.1184$ given by $\pm 3\sigma$. Blue lines present metastability for current
experimental data, red (thick) line corresponds to the stability of the EW vacuum.}
\end{figure}

\newpage

\begin{figure}
\centering
\includegraphics[scale=0.35]{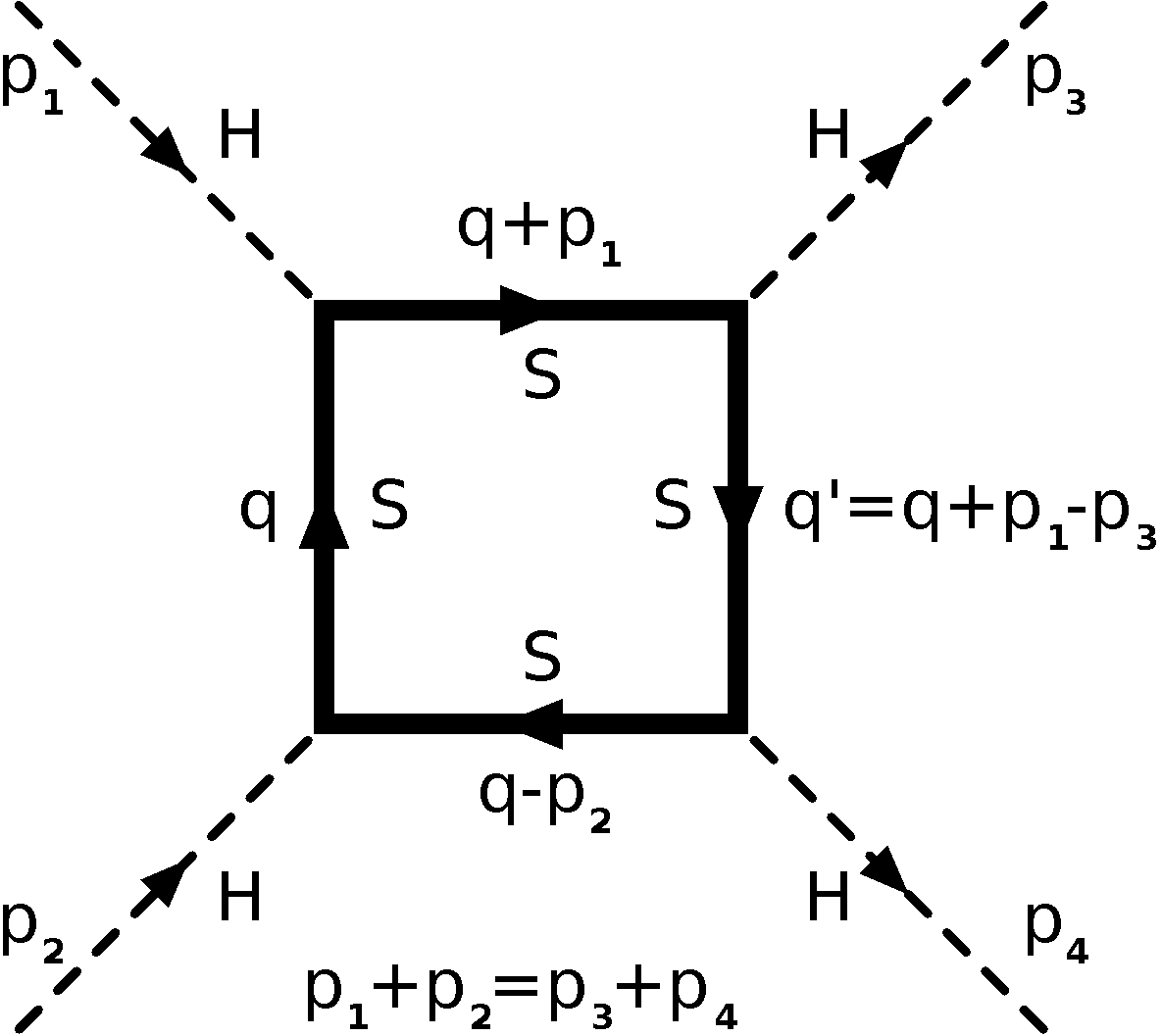}
\caption
{The Feynman diagram corresponding to the main contribution of the $S$ bound
state $6t + 6\bar t$ to the running Higgs self-coupling $\lambda$.}
\end{figure}

\end{document}